\documentclass[prd,twocolumn,showpacs,preprintnumbers,superscriptaddress,floatfix,nofootinbib]{revtex4-1}
\usepackage{amsfonts}
\usepackage{graphicx,,booktabs}
\usepackage{amssymb}
\usepackage{amsmath,txfonts,ulem}
\usepackage{graphicx}
\usepackage{dcolumn}
\usepackage{color}
\usepackage{bm}
\usepackage{ulem}
\usepackage[colorlinks,
            citecolor=blue,
            anchorcolor=green,
            menucolor=orange,
            linkcolor=red,
            filecolor=red,
            runcolor=pink,
            urlcolor=blue,
            frenchlinks=red]{hyperref}

\allowdisplaybreaks[4]

\begin{document}

\title{\boldmath Systematic analysis of the   proton  mass radius based on photoproduction of vector charmoniums }
\author{Xiao-Yun Wang}
\email{xywang@lut.edu.cn}
\affiliation{Department of physics, Lanzhou University of Technology,
Lanzhou 730050, China}
\affiliation{Lanzhou Center for Theoretical Physics, Key Laboratory of Theoretical Physics of Gansu Province, Lanzhou University, Lanzhou, Gansu 730000, China}

\author{Fancong Zeng}
\email{fczeng@yeah.net}
\affiliation{Department of physics, Lanzhou University of Technology,
Lanzhou 730050, China}

\author{Quanjin Wang}
\affiliation{Department of physics, Lanzhou University of Technology,
Lanzhou 730050, China}

\begin{abstract}

In this work, the cross-section of the reaction $\gamma p \rightarrow V(  J / \psi,\psi(2S)) p$ from  the production threshold to medium energy is studied and  systematically  analyzed within  two gluon exchange model.   The obtained numerical results are in agreement with experimental data and other theoretical predictions.
Under the assumption of the scalar form factor of dipole form,
the value of   proton mass radius   is calculated as $0.55\pm 0.09$ fm and  $ 0.77\pm 0.12$  fm  from the fit to the predicted  $J/\psi $  and  $\psi(2S)$  differential cross-section, respectively.
Finally the average value of proton mass radius is estimated to be $\sqrt{\left<R^2_m \right>}=0.67\pm 0.11$ fm.
Moreover, one find that   extracting mass radius from the near-threshold differential cross-section of heavy quarkoniums  is always affected by  large $|t|_{min}$.
These obtained results may provide important theoretical reference for  understanding of nucleon structure and future relevant experiments.
\end{abstract}

\maketitle
\section{INTRODUCTION}

The proton radius is a big inspiration in  understanding the proton structure, and it can be measured by using lepton as probe.
Usually, the proton radius can be estimated by colliding the nucleus with high-energy electrons and observing the angles and energies of these electrons scattered from the nucleus. In the past decade, several groups have given their results \cite{Pohl:2010zza,A1:2010nsl,Antognini:2013txn,Alarcon:2018zbz,Beyer:2017gug,Bezginov:2019mdi,Fleurbaey:2018fih,Mihovilovic:2019jiz,Xiong:2019umf,Zhan:2011ji} and the latest value of the proton charge and  magnetic radius are calculated as $ 0.8409$ fm \cite{ParticleDataGroup:2020ssz} and $ 0.817$ fm \cite{Cui:2021skn}, respectively. In fact, it is theoretically possible to use graviton as a probe to determine the  proton radius, which is usually called proton mass radius. Since the interaction of gravitons and proton scattering is very weak, far beyond the measurement limit of current experiments, it is difficult to directly measure the proton mass radius experimentally.  In recent study, mass radius is described by the scalar gravitational form factors (GFFs) of the energy momentum tensor trace of quantum chromodynamics (QCD) \cite{Kobzarev:1962wt,Pagels:1966zza,Ji:1996nm}. Under the framework of QCD theory, the photoproduction of a quarkonium off the proton is connected to GFFs of the proton, which is sensitive to the proton mass distribution from the  quantum chromodynamics trace anomaly \cite{Wang:2019mza,Ji:1996nm}. Under an assumption of the scalar form factor of dipole form, the proton mass radius can be extracted via the near-threshold photoproduction data of vector quarkoniums \cite{Kharzeev:2021qkd,Fujii:1999xn}.

In recent years, researchers have made relevant calculations and studies on the proton mass radius using vector meson photoproduction data.
In ref. \cite{Guo:2021ibg}, the gluonic contributions to the quantum anomalous energy, mass radius, spin, and mechanical pressure in the proton are studied by analyzing the near-threshold production of  heavy quarkonium, and  proton mass radius is  calculated as 0.68 fm. By studying  the near-threshold differential cross-section of the vector mesons, one work \cite{Wang:2021dis} get the average value of  mass radius is $0.67 \pm 0.03$ fm, according to $\omega$, $\phi$,  $J/\psi$  experiment data \cite{GlueX:2019mkq,LEPS:2005hax,Barth:2003kv}.
Kharzeev calculate the mass radius is $0.55 \pm 0.03$ fm \cite{Kharzeev:2021qkd}   according to GlueX   data of $J/\psi$ photoproduction \cite{GlueX:2019mkq}. One noticed that the proton's mass radius is smaller than its charge radius, which may mean that the mass distribution of protons is tighter than the charge distribution, and different interaction forces correspond to different proton radius.

In the charm quark energy region, due to the mass of $J/\psi$ and $\psi(2S)$ are both below the mass threshold of $D\overline{D}$,  hadron decay via OZI suppression makes their  decay widths very small. The small decay widths give $J/\psi$  and  $ \psi(2S)$  a long life, which is beneficial to observe the relevant physical quantities in the experiment. Therefore, it is of great interest to systematically analyse  the proton mass radius from photoproduction of vector charmoniums.
Currently, the photoproduction of $J/\psi$  has been measured  with an increasing precision over a large energy range \cite{GlueX:2019mkq,ZEUS:2002wfj,Binkley:1981kv,E687:1993hlm,H1:2013okq,ALICE:2014eof,LHCb:2013nqs,ALICE:2018oyo,Amarian:1999pi},
while the measurements of $\psi(2S)$ photoproduction data are very meagre (only some sparse data  exist  near $100$ GeV) \cite{H1:2002yab,LHCb:2018rcm,Hentschinski:2020yfm}. Among them, although GlueX \cite{GlueX:2019mkq} measured the near-threshold photoproduction differential cross-section of $J/\psi$, the corresponding center-of-mass energy was 4.58 GeV, which was several hundred MeV larger than the threshold energy. For $ \psi(2S)$, photoproduction differential cross-section data at threshold is not yet available. Such a situation results in that we can only rely on limited experimental data when extracting the proton mass radius, and it is difficult to systematically analyze and describe the overall influence of the cross-section near the threshold of the vector  charmoniums on the proton mass radius. Therefore, one need to first consider the theoretical calculation and prediction of charmoniums ($J/\psi$ and $ \psi(2S)$) photoproduction with the help of physical models, and then systematically study the mass radius based on the theoretical differential cross-section.

Considering the $\psi(2S)$ and $J/\psi$  have close mass and the same quantum number, it is reasonable to study them with the same physical model and parameters. Since the two-gluon exchanging process may predominate in charmoniums photoproduction \cite{Hentschinski:2020yfm,Zeng:2020coc}, in this work we attempt to extend the study of $J/\psi$ to $ \psi(2S)$ under the framework of the two gluon exchange model, providing data for the systematic extraction of proton mass radius. The paper is organized as follows. The formulas of the two gluon exchange model and the relation between mass radius and differential cross-section  are provided in Sec. \ref{sec:formalism}. Then in Sec. \ref{sec:results}, we show the numerical result on the explanations of the current experimental data of $J/\psi $ photoproduction,   the predictions  of $\psi(2S)$ photoproduction,  and the discussion and results of mass radius $\sqrt{\left<R^2_m \right>}$. A short summary is given in Sec. \ref{sec:summary}.

\section{  FORMALISM}\label{sec:formalism}
 \subsection{two gluon exchange model}
\begin{figure}[h!]
\center
\includegraphics[scale=0.4]{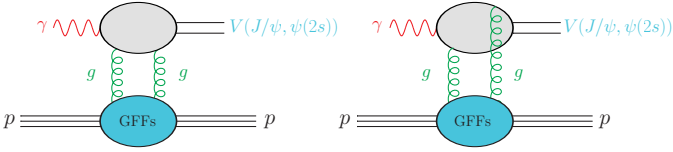}
\caption{The schematic Feynman diagram of the two-gluon exchange   for $V(J/\psi,    \psi(2S)  )$ production. }
\label{fig:TGE-model}
\end{figure}

The two gluon exchange model is based on the photon fluctuation into the quark-antiquark pair ($\gamma \rightarrow c+\bar{c}$)  and the  double gluon exchange between the nucleon state and the quark-antiquark pair. Here the $c\bar{c}$  fluctuation  of the photon treated as a colour dipole. Finally the dipole forms into final vector meson. The picture of the double gluon exchange is illustrated in Fig.\ref{fig:TGE-model}. Here, we describe the process
\begin{equation}
\gamma(q)+p(p) \rightarrow V(q_1)+p(p_1)
\end{equation}
where $V=J/\psi$, $\psi(2S)$.

Due to the hard scale in the heavy quarkonium production, the exclusive vector meson photoproduction amplitude   is given by
\cite{Ryskin:1992ui,Brodsky:1994kf,Ryskin:1995hz}
\begin{equation}\label{eq:amplitudeT}
\begin{split}
\mathcal{T}=&\frac{i \sqrt{2} \pi^{2}}{3} m_{q} \alpha_{s} e_{q} f_{V} F_{2 g}(t) \\
&\left[\frac{x g\left(x, Q_{0}^{2}\right)}{m_{q}^{4}}   +\int_{Q_{0}^{2}}^{+\infty} \frac{d l^{2}}{m_{q}^{2}\left(m_{q}^{2}+l^{2}\right)} \frac{\partial x g\left(x, l^{2}\right)}{\partial l^{2}}\right].
\end{split}
\end{equation}

The amplitude is normalized and $\frac{d \sigma}{d t}=\alpha |\mathcal{T}|^{2}$.
 The $J/\psi$ and $ \psi(2S)$  photoproduction differential cross-section at low momentum transfer are purely diffeactive and  given as \cite{Sibirtsev:2004ca,Zeng:2020coc}
\begin{align}
\label{eq:diff-crosssection}
\frac{d \sigma}{d t}=\frac{\pi^{3} \Gamma^{V}_{e^{+}e^{-}} \alpha_{s}}{6 \alpha m_{q}^{5}}\left[x g\left(x, m_{V}^{2}\right)\right]^{2} \exp (bt),
\end{align}
where $x=m_V^2/W^2$, $W$ is the   center-of-mass energy of the $\gamma p$ collision; $t$ is the momentum transfer.
    $\alpha_{s}=0.5 $ is the strong running  coupling constant \cite{Xu:2020uaa}, $\alpha=1/137$ is the electromagnetic coupling constant, $m_q=1.27$ GeV  is  the mass   of $c$ quark.
$\Gamma^V_{e^+e^-}$ is the radiative decay of vector meson.
In this paper, we set  $\Gamma_{e^+e^-}^{J/\psi}=5.547$ keV and  $\Gamma_{e^+e^-}^{\psi(2S)}=2.33$ keV from PDG \cite{ParticleDataGroup:2018ovx}.
   The   factor $xg\left(x, m_{V}^{2}\right)$   is the gluon distribution function at $Q^2=m_{V}^{2}$, which is parameterized using a simple function form $xg\left(x, m_{V}^{2}\right)=A_0 x^{A_1}(1-x)^{A_2}$ \cite{Pumplin:2002vw}. In this paper, the photoproduction of $J/\psi$ and $\psi(2S)$  will use the same gluon distribution function.

 The last exponential factor in Eq. \ref{eq:diff-crosssection} usually describes the differential cross-section of vector meson at low momentum  transfer.
The exponential slope $b $ for $J/\psi$ has little vary with  $W$, which is given as  \cite{Zeng:2020coc}
\begin{align}\label{bwpsi}
b^{J/\psi}(W)= b_0^{J/\psi}+0.46 \cdot In(W/W_0^{J/\psi})
\end{align}
and fixed the slope $b_0^{J/\psi}= 1.67\pm 0.38$ GeV $^{-2}$  at the energy $W_0^{J/\psi}=4.58$ GeV \cite{GlueX:2019mkq}.
For $b^{\psi(2S)}(W)$, we
use the standard form based on the Regge phenomenology \cite{Hentschinski:2020yfm,Cepila:2019skb}
\begin{align}\label{bwpsi2s}
b^{\psi(2S)}(W)= b_0^{\psi(2S)}+4 \alpha'(0)~  In(W/W_0),~~W_0=90 ~\mathrm{GeV}
\end{align}
 the parameters have been determined from a fit to HERA data that  $b_0^{\psi(2S)}=4.86$, and $\alpha'(0) =0.151$ \cite{Hentschinski:2020yfm}.

The total cross-section is obtained by integrating the differential cross-section (Eq. \ref{eq:diff-crosssection})
over the allowed kinematical range from $t_{min}$ to $ t_{max}$,
which  can be written as
\begin{align}
\sigma=\int_{t_{min}}^{t_{max}}  dt  ~\left(\frac{d \sigma}{d t}  \right)
\end{align}
 here, the limiting values $t_{min}$ and  $t_{max}$ are
\begin{align}
 t_{max}\left(t_{min}\right)=\left[\frac{m_{1}^{2}-m_{3}^{2}-m_{2}^{2}+m_{4}^{2}}{2 W}\right]^{2}-\left(p_{1 \mathrm{cm}} \mp p_{3 \mathrm{cm}}\right)^{2}
\end{align}
The center-of-mass energies and momenta of the incoming  photon and vector meson are
\begin{align}
p_{i \mathrm{cm}}=\sqrt{E_{i \mathrm{cm}}^{2}-m_{i}^{2}}~~ (i=1,3),
\end{align}
\begin{align}
E_{1 \mathrm{cm}}= \frac{ W^2+m_{1}^{2}-m_{2}^{2}  }{2 W},~ \mathrm{and} ~
 E_{3 \mathrm{cm}}=\frac{W^2+m_{3}^{2}-m_{4}^{2}   }{ 2 W}.
\end{align}

In addition, one noticed that in refs. \cite{Kharzeev:1995ij,Kharzeev:1998bz,Gryniuk:2016mpk}, the real part of the $J/\psi-p$ scattering amplitude is of great success in vector meson photoproduction around the threshold, which indicate that the real part of the
scattering amplitude is in general important at low energies. Considering the trace terms in the OPE can no longer be neglected at low energies, more theoretical research on the two gluon exchange model and related physical mechanisms is still needed.

\subsection{GFFs and mass radius}
The mass radius can be defined in terms of the scalar gravitational form factors $G(t)$,
 the definition of the mass radius is given by \cite{Miller:2018ybm,Kharzeev:2021qkd}
\begin{align}
\left<R^2_m \right> \equiv \frac{6}{M} \left.\frac{dG(t)}{dt}\right|_{t=0}
\end{align}
with $G(0)=M$.
And the scalar gravitational form factor is defined as
\begin{align}\label{eq:G}
G(t)=\frac{M}{(1-t/m_s^2)^2}
\end{align}
in which $m_s$ is a free parameter.
According to the definition, the mass radius is
connected to the dipole parameter $m_s$ as
\begin{align}
\label{eq:R}
\left<R^2_m \right>=\frac{12}{m_s^2}
\end{align}
The differential cross-section of the photoproduction of the quarkonium can be described with the GFFs, which is written as \cite{Kharzeev:2021qkd,Mamo:2019mka,Hatta:2019lxo,Frankfurt:2002ka}
\begin{align}\label{eq:g2}
 \frac{d \sigma}{dt} \propto G^2(t)
\end{align}

Therefore, it is an effective way  to extract proton mass radius by learning the near-threshold photoproduction data of charmoniums.

\section{Results and discussion}
\label{sec:results}

\subsection{cross section of charmoniums photoproduction}

\begin{figure}[htbp]
\begin{center}
\includegraphics[scale=0.4]{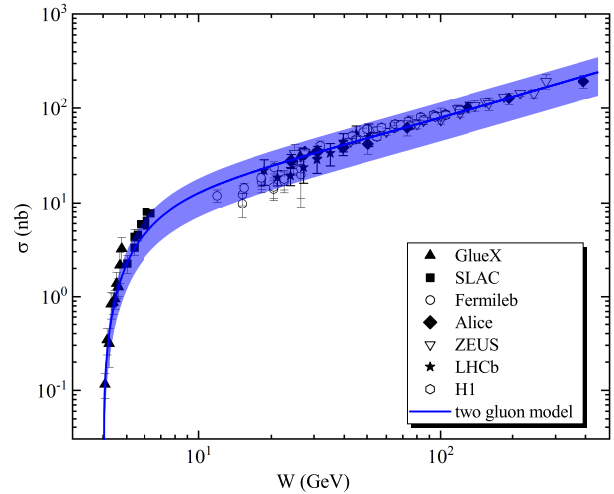}
\caption{The total cross-section of $\gamma p \rightarrow J / \psi p$ as a function of $W$ in two gluon exchange model. The  band  reflect the error bar of the $A_0$.   References of  data  can be found in \cite{GlueX:2019mkq,ZEUS:2002wfj,Binkley:1981kv,E687:1993hlm,H1:2013okq,ALICE:2014eof,LHCb:2013nqs,ALICE:2018oyo,Amarian:1999pi}.}
\label{fig:JPsiTotalCrosssection}
\end{center}
\end{figure}

\begin{figure}[htbp]
\begin{center}
\includegraphics[scale=0.4]{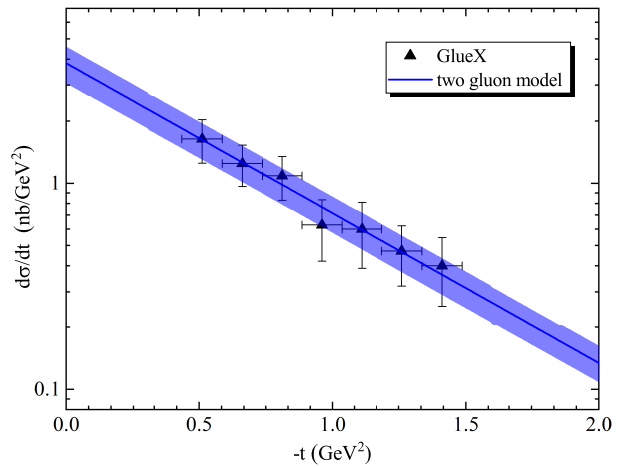}
\caption{The differential cross-section of $\gamma p \rightarrow J / \psi p$ as a function of $-t$ in two gluon exchange model. Reference of  data  can be found in  \cite{GlueX:2019mkq}.}
\label{fig:differental}
\end{center}
\end{figure}

The parametrization gluon distribution function $xg(x,m_{V}^2)=A_0 x^{A_1} (1-x)^{A_2}$
is introduced and used in two gluon exchange model discussed in  the above section. The free parameters $A_0, A_1, A_2$
then are fixed by a global analysis of both the total cross-section data below medium energy  (c.m. energy  near  $ 400$ GeV)  \cite{GlueX:2019mkq,ZEUS:2002wfj,Binkley:1981kv,E687:1993hlm,H1:2013okq,ALICE:2014eof,LHCb:2013nqs,ALICE:2018oyo} and   the near-threshold ($W=4.58$ GeV) differential cross-section data  \cite{GlueX:2019mkq} of $J / \psi$. The obtained parameters of gluon distribution are listed in Table \ref{tab:GluonParameters}.
The  total cross-section of $\gamma p \rightarrow J / \psi p$ as a function of c.m. energy $W$ is shown in Fig.\ref{fig:JPsiTotalCrosssection}, compared to several experimental data \cite{GlueX:2019mkq,ZEUS:2002wfj,Binkley:1981kv,E687:1993hlm,H1:2013okq,ALICE:2014eof,LHCb:2013nqs,ALICE:2018oyo,Amarian:1999pi}.
The comparison between the differential cross-section
and the experimental measurements is manifested in Fig.\ref{fig:differental},
exhibiting a good agreement. Finally one notice that the $\chi^{2}/d.o.f.$ of our global fit is calculated to be 2.066, which indicates that the parameters obtained from gluon distribution is applicable for $J/\psi$ photoproduction.

\begin{table}
\footnotesize
\centering
\caption{The fitted values of the parameters $A_0, A_1, A_2$ describing the gluon distribution function $xg(x)$ according to $ J / \psi$ photoproduction.
and the reduced $\chi^{2}/d.o.f.$ }
\begin{tabular}{ccccc}
\hline\hline\noalign{\smallskip}
  $A_0$ & $A_1$ & $ A_2$ &   $\chi^{2}/d.o.f.$    \\
\noalign{\smallskip}\hline\noalign{\smallskip}
 ~~~~ $0.228\pm 0.045$~~~~ & ~~~~$-0.218\pm 0.006$~~~~ & ~~~~$1.221\pm0.055$~~~~ & ~~~~ 2.066~~~~ \\
\noalign{\smallskip}
\hline
\end{tabular}
\label{tab:GluonParameters}
\end{table}

\begin{figure}[htbp]
\begin{center}
\includegraphics[scale=0.4]{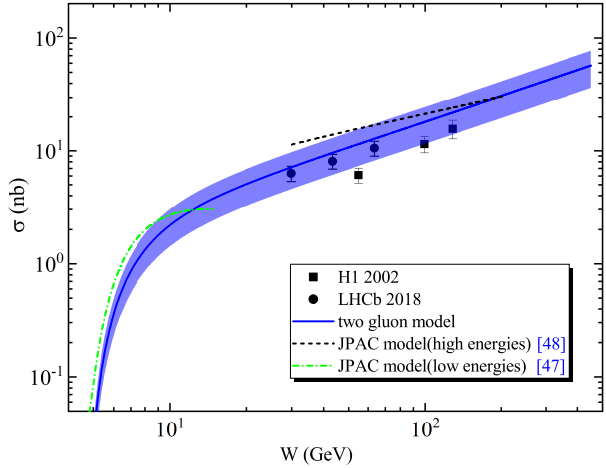}
\caption{The total cross-section of $\gamma p \rightarrow  \psi(2S) p$ as a function of $W$.
Blue solid curve is abtained from two gluon exchange model.  Black dotted  curve and green dot-dashed curve are abtained from the prediction of JPAC  in  high energies and low energies respectively.
    References of  data  can be found in \cite{H1:2002yab,LHCb:2018rcm,Hentschinski:2020yfm}.}
\label{fig:2s}
\end{center}
\end{figure}

\begin{figure}[htbp]
\begin{center}
\includegraphics[scale=0.4]{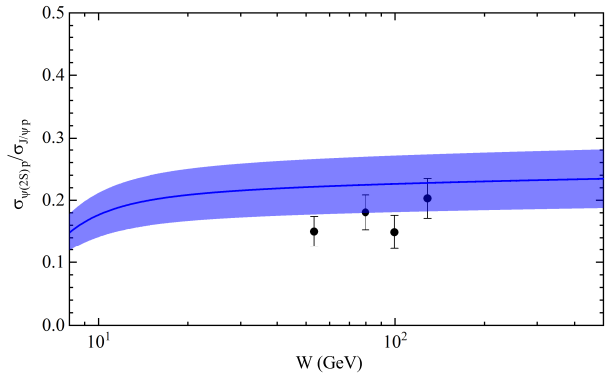}
\caption{The result of   $J/\psi p$  and  $\psi(2S) p$ total cross-section ratio $R$ as a function of $W$. Reference of  data  can be found in \cite{H1:2002yab}.}
\label{fig:ratio}
\end{center}
\end{figure}

\begin{figure}[htbp]
\begin{center}
\includegraphics[scale=0.4]{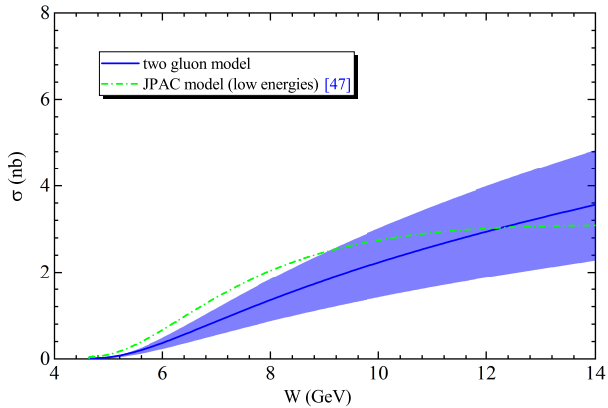}
\caption{The total cross-section of $\gamma p \rightarrow  \psi(2S) p$ as a function of $W$. The curves have the same meaning as in Fig.\ref{fig:2s}. }\label{fig:small-w}
\end{center}
\end{figure}

Using the parameters in Tab. \ref{tab:GluonParameters}, one calculated the photoproduction cross-section of $\psi(2S)$, as shown by the blue solid curve in Fig.\ref{fig:2s}. It is shown that the theoretical calculations are in good agreement with the experimental datas \cite{H1:2002yab,LHCb:2018rcm,Hentschinski:2020yfm} of $\psi(2S)$ photoproduction. Moreover, the ratio of $\psi(2S)$ to $J/\psi$  total cross-section $R=\sigma_{\psi(2S)p}/\sigma_{J/\psi p}$ is estimated  as a function of $W$  in Fig.\ref{fig:ratio}, which also agrees well with the experimental data \cite{H1:2002yab}. The above results indicate that it is suitable and feasible to extend the study from $J/\psi$ to $\psi(2S)$ production under the framework of the two gluon exchange model. Moreover, it is worth mentioning that two effective Pomeron models are considered  to research the $\psi(2S)$ photoproduction in low and high energy regions by JPAC \cite{Winney:2019edt,Albaladejo:2020tzt}. We compared our results with  JPAC models, which are shown in  Fig.\ref{fig:2s} and \ref{fig:small-w}. The prediction of JPAC models are basically consistent with our results at the threshold, while it is larger than experimental data in higher energy regions.

\subsection{proton mass radius}

\begin{figure}[htbp]
\begin{center}
\includegraphics[scale=0.4]{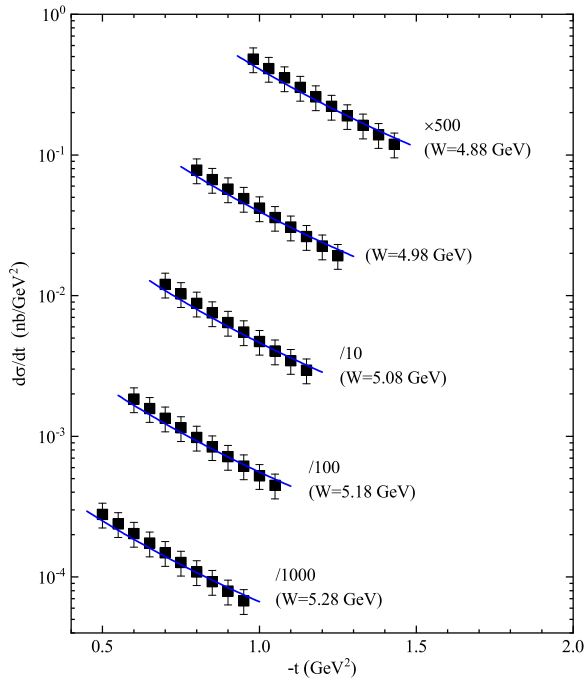}
\caption{
The differential cross-section of $\psi(2S)$ in  Eq. \ref{eq:g2}  shown in  blue solid curve. Black squares shows the predicted differential   cross-section of $\gamma p \rightarrow   \psi(2S) p$ as a function of $-t$.}
\label{fig:diffent2s}
\end{center}
\end{figure}

\begin{figure}[htbp]
\begin{center}
\includegraphics[scale=0.4]{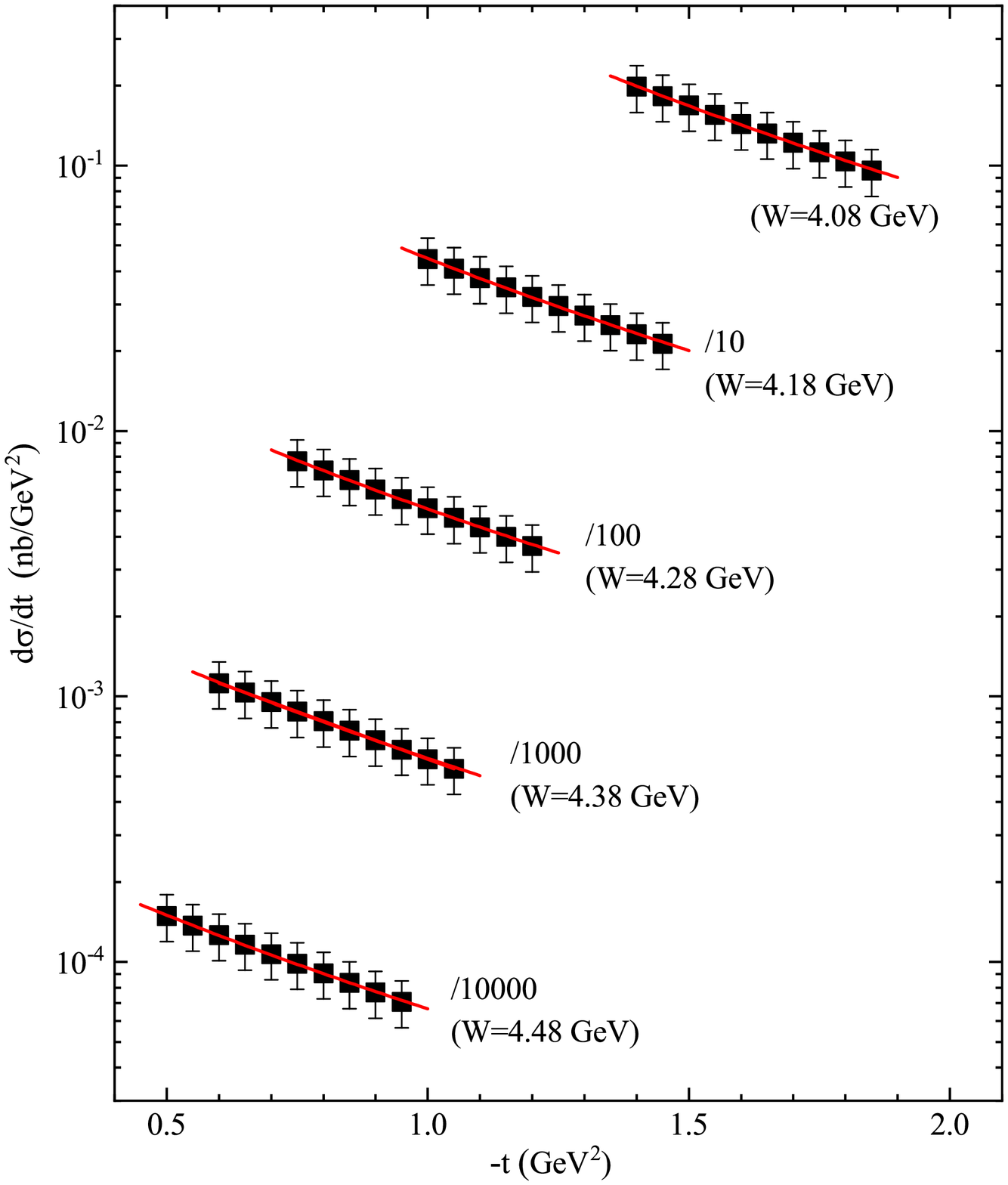}
\caption{
The differential cross-section of $J/\psi $ in Eq. \ref{eq:g2}  shown in red solid curve. Black squares   shows the predicted differential   cross-section of $\gamma p \rightarrow   J/\psi  p$ as a function of $-t$.}
\label{fig:diffentpsi}
\end{center}
\end{figure}

\begin{figure}[htbp]
\begin{center}
\includegraphics[scale=0.42]{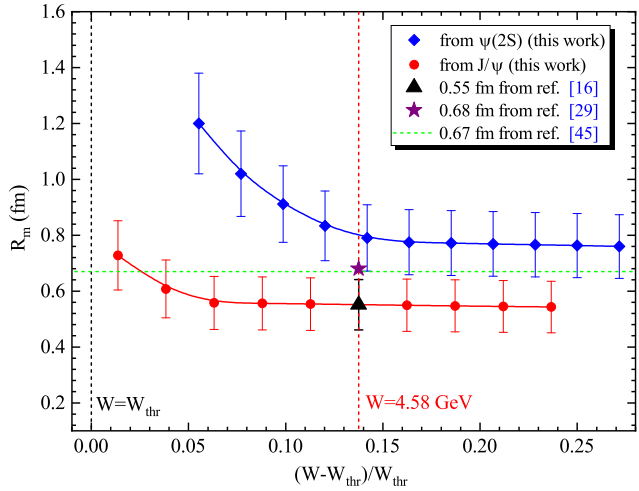}
\caption{ The extracted mass radius at different  c.m. energy from predicted $\psi(2S)$ photoproduction (blue squares) and $J/\psi$ photoproduction (red circles). The green line shows the  result of ref.\cite{Wang:2021dis}. The purple pentagram shows the result of X. Ji\cite{Guo:2021ibg}. The black triangle shows the result of  Kharzeev\cite{Kharzeev:2021qkd}. }
\label{fig:last}
\end{center}
\end{figure}

Under the assumption of the scalar form factor of dipole form, the proton mass radius can be  extracted from the near-threshold differential cross-section of vector charmoniums.
In this work, the differential   cross-section of $\gamma p \rightarrow   \psi(2S) p$
are predicted by two gluon exchange model. We choose the c.m. energy $W\in(4.88,5.28)$ interval of 0.1 GeV and a range of $\Delta t= 0.5$ GeV$^2$  starting with $|t|_{min}$, which are shown in the black squares  in Fig.\ref{fig:diffent2s}.
The blue solid curve in  Fig.\ref{fig:diffent2s} is the fitted scalar gravitational form factor of the dipole parametrization in  Eq. \ref{eq:g2}.
Finally, we get the value of proton mass radius from  Eq. \ref{eq:R}  and \ref{eq:g2}, which  are  the first five blue squares in  Fig.\ref{fig:last}. Here, the small $x$-axis $(W-W_{thr})/W_{thr}$ represents the c.m. energy  close to the threshold energy $W_{thr}$.
 For $J/\psi$ mesons,
the  numerical results obtained from similar methods can be seen in Fig.\ref{fig:diffentpsi} and the  first five red circles in Fig.\ref{fig:last}.

\begin{figure}[htbp]
\begin{center}
\includegraphics[scale=0.4]{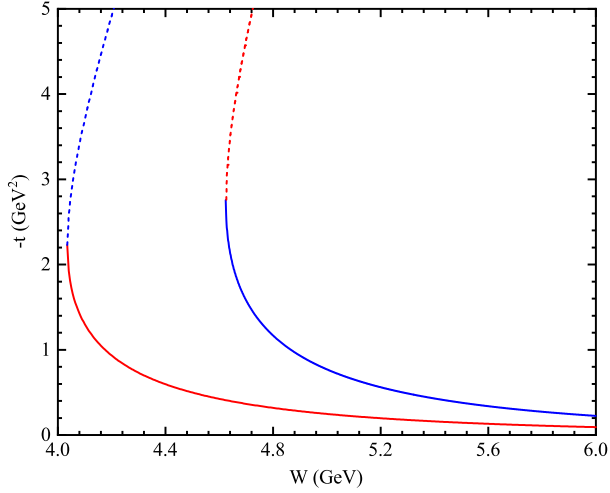}
\caption{the limiting values $t_{min}$ and  $t_{max}$  as a function of $W$. Bule solid curve  shows  $|t|_{min}$ of $\psi(2S)p$ photoproduction,   Red solid curve shows  $|t|_{min}$ of $J/\psi p$ photoproduction.  }
\label{fig:tmin}
\end{center}
\end{figure}

Note that, the proton mass radius mainly depends on the slope of the vector meson threshold photoproduction cross-section, and is hardly affected by the cross-section size. However, one found that although the slope $b$ at the threshold changes very slowly (which can be verified by Eq. \ref{bwpsi} and \ref{bwpsi2s}), the extracted mass radius changes sharply at the threshold, as shown in Fig.\ref{fig:last}.
Actually, the mainly reason is the rapidly varies of $|t|_{min}$ near threshold. Accordingly, one can give the invariant momentum transfer $|t|$  at the threshold  equals
\begin{align}
|t|_{thr}=|t|_{min}(W_{thr})=|t|_{max}(W_{thr})=\frac{M_N M_V^2}{M_N+ M_V}
\end{align}
By analyzing the formula, it can be obtained that when the mass of the vector meson is heavier, the corresponding $|t|_{thr}$ is also larger, which eventually leads to $|t|_{min}$ have  more rapid  variation near threshold as shown in Fig.\ref{fig:tmin}. This indicates that  $|t|_{min}$ has greater effect on the heavier vector mesons. The above situation requires us to pay special attention when extracting the proton mass radius, and to perform overall analysis and extraction to avoid single-point dependence.
For some light vector mesons (such as $ \phi$ mesons, etc.) photoproduction, since the change of $|t|_{min}$ at the threshold is relatively gentle, it is beneficial to the experimental measurement of the near-threshold cross-section and the extraction of the mass radius.

As shown in Fig.\ref{fig:last},
the  influence of big $|t|_{min}$ on extracting mass radius in $\psi(2S)p$ and $J/\psi p$ photoproduction can be eliminated with the increase of c.m. energy. Since the extracted mass radius gradually tends to a stable value with the increase of energy, we consider selecting an energy interval closest to the threshold, and determine the stable value of the mass radius in this energy interval as the required physical quantity. Thus the average value of dipole parameter $m_s=0.88\pm 0.11$ GeV and mass radius is calculated as $ 0.77\pm 0.12$ fm from the $\psi(2S)p$ differential cross-section with $W\in(5.28,5.88)$.
The mass radius is calculated as $ 0.55\pm 0.09$ fm from  $J/\psi p$ photoproduction with $W\in(4.38,5.98)$.
Finally the average value of the proton mass radius is calculated to be $\sqrt{\left<R^2_m \right>}=0.67\pm 0.11$ fm.

By studying  the near-threshold   differential cross-section of the vector mesons $\omega$, $\phi$,  $J/\psi$, one work
 \cite{Wang:2021dis}
get the average value of the proton mass radius is   $0.67 \pm 0.03$ fm (the green dotted curve in Fig.\ref{fig:last}).
Kharzeev calculate the    mass radius $0.55 \pm 0.03$ fm      (the black triangle in Fig.\ref{fig:last}) \cite{Kharzeev:2021qkd}   according to GlueX   data of $J/\psi$ photoproduction\cite{GlueX:2019mkq},
while X. Ji calculates that the proton mass radius is 0.68 fm (the purple pentagram Fig.\ref{fig:last}) \cite{Guo:2021ibg}.
In fact, if the error bar is considered, our results    are   in agreement with    the above theoretical predictions.

\section{Summary}
\label{sec:summary}

We have reproduced the total and differential cross-section
of the reaction $\gamma p \rightarrow J / \psi p$ from  the production threshold to medium energy ($W$ near  $ 400$ GeV) with the two gluon exchange model encountering a parameterized gluon distribution function. By fitting the experimental data, one get the parameters of the gluon distribution function and the result show that the two gluon exchange model depicts well both the differential and total cross-section of $J/\psi$ in a wide energy range.
Subsequently, one estimate  the photoproduction of $\psi(2S)$ meson using the gluon distribution determined by  $J/\psi$ photoproduction. The comparison between our theoretical prediction and $\psi(2S)$ experimental data is well.
Naturally, the proton mass radius is extracted by using the predicted differential cross-section of charmoniums ( $J/\psi$ and $\psi(2S)$). After system analysis,  the average value of the proton mass radius is estimated to be $0.67\pm 0.11$ fm. This value is smaller than proton charge and magnetic radius, but is basically close to the mass radius value given by other theoretical groups \cite{Wang:2021dis,Guo:2021ibg,Kharzeev:2021qkd}.
We also find that  extracting mass radius from the near-threshold differential cross-section is always affected by  large $|t|_{min}$.  This requires us to take special care in extracting  mass radius through the near-threshold cross-section of heavy vector quarkoniums photoproduction. Of course, more accurate experimental measurement data for the photo/electro-production of charmoniums is still needed, which can not only be realized in the JLab experiment \cite{GlueX:2019mkq}, but also within the capabilities of EicC and US-EIC facility \cite{Anderle:2021wcy,Accardi:2012qut}.
Moreover, the results of  charmoniums   photoproduction will provide an important theoretical reference  for   Ultraperipheral collisions (UPCs) \cite{Klein:2016yzr,Klein:2019avl,Wang:2020stx,Xie:2020wfe}.
Therefore, it would also be interesting to investigate the production of charmoniums in $ep$ and $pA$ collisions according to the actual situation of EIC and UPCs experiments.

\section{Acknowledgments}

X.-Y. Wang would like to acknowledge Dr. Daniel Winney for useful discussion about the JPAC model.
This project is supported by the National Natural Science Foundation of China (Grant Nos. 12065014 and 12047501),
and by the West Light Foundation of The Chinese Academy of Sciences, Grant No. 21JR7RA201.

\end{document}